\documentclass[aps,prc,a4paper,twocolumn,nofootinbib]{revtex4}
\usepackage{graphicx}
\usepackage{bm}
\usepackage{amsmath}
\usepackage{amsfonts}

\begin{document}
\title{
A new formulation to calculate general HFB matrix elements through Pfaffian
}

\author{Takahiro Mizusaki}
\affiliation{Institute of Natural Sciences, Senshu University, 
3-8-1 Kanda-Jinbocho, Chiyoda-ku,Tokyo 101-8425, Japan}

\author{Makito Oi}
\affiliation{Institute of Natural Sciences, Senshu University, 
3-8-1 Kanda-Jinbocho, Chiyoda-ku,Tokyo 101-8425, Japan}
\email{m.oi@isc.senshu-u.ac.jp}


\begin{abstract}
A new formula is presented for the calculation of matrix elements 
between multi-quasiparticle Hartree-Fock-Bogoliubov (HFB) states.
The formula is expressed in terms of the Pfaffian, and is derived 
by using the Fermion coherent states with Grassmann numbers.
It turns out that the formula corresponds to an extension of generalized Wick's theorem
and simplifies the combinatorial complexity resulting from practical 
applications of generalized Wick's theorem by unifying the transition density and 
the transition pairing tensor in the HFB theory.
The resultant formula is simpler and more compact than the traditional description of 
matrix elements of general many-body operators.
In addition, through the derivation of our new formula, 
we found that the Pfaffian version of the Lewis Carroll formula corresponds to the relation 
conjectrured by Balian and Brezin for the HFB theory in 1969.
\end{abstract}

\maketitle

\section{Introduction} 

A Fermion coherent state with Grassmann numbers has been useful not only to particle physics, 
but also to quantum many-body physics \cite{NO85}.
For instance, traditional perturbation theories both for zero and 
finite temperature quantum systems can be formulated by means of the functional integral
with the Fermion coherent state and Grassmann numbers\cite{NO85}. 
With this approach, we can obtain a new insight about Green's functions, 
partition functions in statistical mechanics,
and many-body matrix elements (overlaps for many-body states), and so on.

Such a Fermion coherent state with Grassmann numbers 
was recently applied to nuclear structure physics by Robledo \cite{Rob09}. 
His discovery is that the Thouless form of  
Hartree-Fock-Bogoliubov (HFB) state can be mapped to the Fermion coherent state
with the help of Grassmann numbers. 
It sheds a new light on hitherto the established formulation and the long-standing unresolved issues 
in nuclear many-body systems.

In nuclear structure physics, the HFB method has played a central role in describing interacting nucleons 
through a short-range effective force (e.g., pairing interaction) as well as a long-range force 
(e.g., quadrupole interaction) \cite{RS80}. With various constrained conditions, 
the HFB has succeeded in giving a good mean-field description for various many-body states.
However, in an attempt to go beyond the mean-field description, especially 
in the case of three-dimensional angular momentum projection, there has been a difficulty 
originating from the long-standing problem in the phase determination of norm-overlap kernels 
through the Onishi formula \cite{OY66}.
This problem has been thoroughly investigated by many authors,  
for instance, in Refs. \cite{OY66,HHR82,NW83,ETY99,OT05}. 
Many of them relied upon the analytic continuity approach
for the phase determination, which can be carried out with the Onishi formula\cite{OY66}.
It was demonstrated recently that this approach works  even for high-spin states 
in the presence of many nodal lines \cite{OMlimbo}. 

On the contrary to these pursuits, Robledo was successful in  deriving a new formula
for the HFB norm-overlap kernel in terms of the Pfaffian. The formula contains no ambiguity in phase.
This achievement was made by means of the Fermion coherent states with Grassmann numbers \cite{Rob09}.
By his formula, the perennial problem has been finally solved in a mathematically elegant fashion.
We extended his methodology to find another formula, which is applicable both to 
even- and odd-mass nucleon systems \cite{OM11}. 

A further extension should be to derive a formula applied to multi-quasiparticle states.
A standard approach to evaluate matrix elements between multi-quasiparticle states 
is to use generalized Wick's theorem. By this theorem, evaluation
of the matrix elements between multi-quasiparticle states is reduced to a combinatorial 
problem with contractions, which correspond to the density matrix and the pairing tensor
in the HFB theory 
(the transition density and the transition pairing tensor in the case of projected HFB).
However, due to the combinatorial nature, the number of combinations of the contractions rapidly increases 
as we consider more complicated multi-quasiparticle states.
Therefore, it is important to derive a systematic, compact and closed form for the 
formula in order to calculate such matrix elements effectively.
Perez-Martin and Robledo derived a formula based on the finite temperature 
formalism\cite{Rob07}. 
Recently, different formulae expressed in terms of  the Pfaffian have been proposed
by Bertsch and Robledo\cite{BR11}, and by Avez and Bender\cite{AB12}, although 
the latter does not correspond to generalized Wick's theorem.

In this paper, we also present a new compact and closed form (but more compact than the above former two) 
with the Pfaffian by means of the Fermion coherent state 
with Grassmann numbers. In Ref.\cite{OM11}, we have already presented a 
Pfaffian formula to calculate the norm-overlap kernels of one-quasiparticle states. 
This formula is extended to multi-quasiparticle cases in the present work,
by introducing a generating function expressed by Grassmann numbers. 
It turns out that the formula corresponds to an extension of  generalized Wick's theorem.
By comparing our formula and the one by Avez and Bender\cite{AB12}, we can give an 
affirmative answer to the Balian-Brezin conjecture \cite{BB69}. 
It is surprising to find that a usefulness of the Pfaffian in the HFB theory has been already
suggested in the their paper in 1969.
 

The present paper is organized as follows. Section II is devoted to an explanation of the 
general mathematical structure of matrix elements of many-body operators.
Section III is for the Grassmann representation of the matrix elements. 
Section IV is a brief review on the Grassmann integrations. Section V 
and VI present a derivation of a Pfaffian form of the considered matrix elements, 
and a discussion concerning generalized Wick's theorem. 
In Section VII, a discussion is given on the Balian-Bresin conjecture. 
Section VIII is the summary of the present paper.

Because the references necessary for the present paper cover a wide range of fields in phyiscs and mathematics,
such as particle physics, mathematical physics, combinatorics in pure mathematics and so on, 
it is convenient to structure this paper in a self-contained manner.


\section{Matrix element for multi-quasiparticle states}
In this paper, the central representation of the HFB wave function is 
the Thouless form\cite{RS80}, which is presented by
\begin{equation}
  |\Phi^{(p)}\rangle = \mathcal{N}\exp\left(\frac{1}{2}\sum_{i<j}^M Z_{ij}^{(p)}c_i^{\dag}c_j^{\dag}\right)|0\rangle,
  \label{Thouless}
\end{equation}
where $p$ takes 0 or 1, and $\mathcal{N}$ is a normalization factor.
The creation (annihilation) operator of a single-particle orbit $i$
is expressed as $c_i^{\dagger}$ ($c_i$). 
The total dimension of the single-particle model space is $M$. 
$|0\rangle$ is the vacuum state to this basis, that is, $c_i|0\rangle=0$.
Anti-symmetric matrix  $Z$ is expressed in terms of the Bogoliubov transformation 
(or the so-called $U$ and $V$ matrices), as
\begin{equation}
  Z^{(p)} = (V^{(p)}(U^{(p)})^{-1})^*.
  \label{ZVU}
\end{equation}

An $m$-quasiparticle  HFB state is given by 
\begin{equation}
 |\Phi^{(p)}_{{k_1},\cdots,{k_m}}\rangle 
  = \beta_{k_1}^{(p)\dagger}  \cdots \beta_{k_m}^{(p) \dagger} |\Phi^{(p)}\rangle, 
  \label{Quasiparticle}
\end{equation}
where $k_1, k_2, \cdots, k_m$ are indices for quasiparticle states, and 
$\beta$'s ($\beta^{\dag}$'s) are quasiparticle annihilation (creation) operators, 
which are defined with the Bogoliubov transformation, as
\begin{equation}
  \beta_{k}^{(p)\dagger} = U^{(p)}_{i,k}c{_i}^{\dag} + V^{(p)}_{i,k}c_i.
  \label{UV}
\end{equation}

An overlap between different $m_0$- and $m_1$-quasiparticle states
\begin{eqnarray}
 && \langle \Phi^{(0)}_{{k_1},\cdots,{i_{m_0}}} |\Phi^{(1)}_{{k'_1},\cdots,{k'_{m_1}}}\rangle  \nonumber \\
 &=& 
 \langle \Phi^{(0)}|\beta_{k_1}^{(0)} \cdots \beta_{k_{m_0}} ^{(0)}
 \beta_{k'_1}^{(1) \dagger} \cdots \beta_{k'_{m_1}}^{(1) \dagger}|\Phi^{(1)}\rangle, 
 \label{matele_qp}
\end{eqnarray}
is decomposed to a linear combination of the following matrix elements
with $2m=m_0 + m_1$,
\begin{eqnarray}
&&  \mathcal{M}_I = \mathcal{M}_{ \{ {i_1},\cdots,{i_{2m}} \} }  \nonumber \\
&=& 
\langle \Phi^{(0)}|c_{j_1}\cdots c_{j_k}c^{\dagger}_{j'_{1}}\cdots c^{\dagger}_{j'_{2m-k}}
|\Phi^{(1)}\rangle  \nonumber \\
&=& 
\langle \Phi^{(0)}|d_{i_{2m}}\cdots d_{i_{1}}|\Phi^{(1)}\rangle, 
\label{m_qp}
\end{eqnarray}
where $1\le j_k \le M$, $1\le j_{k}' \le M$ and $0\le k \le 2m$. 
Indices $j_k$ and $j_k'$ are relabelled with $i$'s as $i_1 < i_2 < \cdots < i_{2m}$.
Hereafter, we denote this set of indices $\{i_k; k=1,2,3,\cdots, 2m\}$ as $I$.
$I$ is a subset of $[2M]$, that is, 
$I\equiv \{i_1, i_2, \cdots ,i_{2m}\} \subset \{ 1, 2, 3, \cdots, 2M\} \equiv [2M]$.
Here we use a compact notation for a group of indices by following the 
Ref.\cite{IW00}. This notation gives us a good view for a formula with complicated indices. 

It is convenient to introduce a new expression $d_i$ for
\begin{eqnarray}
d_i= \left\{ \begin{array}{ll}
c{_i}^{\dag} & (i=1,\cdots, M)   \\ 
c_{\bar{i}} & (i=1+M,\cdots, 2M)\\
\end{array} \right.
\label{d_op}
\end{eqnarray} 
where $\bar{i} \equiv 2M+1-i$.
By this definition,
an ordering concerning the $d$ operators in Eq.(\ref{m_qp}) can be systematically handled.
As a result, we can deal with products consisting of the arbitrary number of $c$ and $c^{\dagger}$ 
in a unified way.
It should be noted that in the ordering of the creation-annihilation operators in Eq.(\ref{m_qp}), 
all the annihilation operators $c$ are placed in the left of the creation operators $c^\dagger$.
This arrangement needs to be considered for the convenience of the mapping to Grassmann numbers. 
 
\section{Mapping by Fermion Coherent state}
By introducing the Fermion coherent state\cite{NO85}
\begin{equation}
  |\bm{\xi}\rangle = \text{e}^{-\sum_{i}\xi_{i}c_{i}^{\dag}}|0\rangle,
\end{equation}
we map the creation and annihilation operators to Grassmann numbers $\xi^*$ and $\xi$ respectively,
which follow the anticommutation rules
\begin{eqnarray}
\xi_i\xi_j + \xi_j\xi_i &=& 0,\\
\xi_i^*\xi_j^* + \xi_j^*\xi_i^* &=& 0, \\
\xi_i\xi_j^* + \xi_j^*\xi_i &=& 0. 
\end{eqnarray}
By definition, the coherent state is introduced as an eigenstate of the annihilation operator,
\begin{equation}
  c_i|\bm{\xi}\rangle = \xi_i|\bm{\xi}\rangle,
  \label{c-map}
\end{equation}
where $\xi_i$ needs to be a Grassmann number due to the anticommutation nature of the annihilation operators.
The conjugate relation is 
\begin{equation}
  \langle \bm{\xi} |c_i^{\dagger} = \langle \bm{\xi} |\xi_i^*,
  \label{cdag-map}
\end{equation}
where $\xi_i^*$ is a conjugate variable to $\xi_i$.
An expectation value of the $d$ operator in Eq.(\ref{d_op}) is mapped to 
a Grassmann number as,
\begin{equation}
  \langle \bm{\xi}|d_i|\bm{\xi}\rangle = \bar{\xi}_i,
  \label{d-map}
\end{equation}
where  $\bar{\xi}_i$ corresponds to  $\xi_i^*$ for $i=1,\cdots, M$ 
and $\xi_{\bar{i}}$ for $i=1+M,\cdots, 2M$.
A vector $d$ consisting of operators $c$ and $c^{\dagger}$ as 
\begin{equation}
d=(c_1^{\dagger},c_2^{\dagger},\cdots, c_{M}^{\dagger},
c_M,c_{M-1},\cdots,c_1)
\end{equation}
can be mapped to a Grassmann vector $\bar{\xi}$ as 
\begin{equation}
 \bar{\xi}^t \equiv (\xi_1^*\,\xi_2^*,\cdots,\xi_M^*,
  \xi_M,\xi_{M-1},\cdots,\xi_1).
\label{orderofxi}
\end{equation}

Next, we map the general matrix element $\mathcal{M}_I$ in Eq.(\ref{m_qp})
to Grassmann numbers by inserting the completeness
\begin{equation}
\int \prod_{\alpha}d\xi_\alpha^* d\xi_\alpha e^{-\Sigma_j \xi_j^* \xi_j}
|\bm{\xi}\rangle  \langle \bm{\xi}|=1,
\end{equation}
between $c_{j_k}$ and $c^{\dagger}_{j'_{1}}$ in  Eq.(\ref{m_qp}).
By noting that $\langle 0|\xi\rangle=1$, we can obtain 
an expression of the matrix element $\mathcal{M}_I$ in Eq.(\ref{m_qp}) 
in terms of  the following Grassmann integral 
\begin{equation}
 \mathcal{M}_I=
 \int \mathcal{D}\bar{\xi}
 \exp\left({\frac{1}{2}\bar{\xi^t}\mathbb{X}\bar{\xi}}\right)
 \bar{\xi}_{i_{2m}} \cdots \bar{\xi}_{i_{1}},
  \label{qpmat_grassmann}
\end{equation}
where $\mathcal{D}\bar{\xi}=d\bar{\xi}_{2M}\cdots d\bar{\xi}_{1}
=(-1)^M \prod_{\alpha}d\xi_\alpha^* d\xi_\alpha$.
Hereafter $M$ is assumed to be even.
The matrix element $\mathcal{M}_I$ can be shown by the Grassmann integral in terms of only $\bar{\xi}$. 
This fact directly leads to a Pfaffian form as will be discussed in a next section.

In the zero quasiparticle case, the corresponding matrix element is just a norm-overlap kernel
of the original HFB state Eq.(\ref{Thouless}).
Robledo \cite{Rob09} was the first to give a proof that the norm-overlap is represented 
in terms of  the Pfaffian, that is,
\begin{eqnarray}
\langle \Psi^{(0)} | \Psi^{(1)}\rangle 
&=&\int \mathcal{D}\bar{\xi}
\exp\left({\frac{1}{2}\bar{\xi^t}\mathbb{X}\bar{\xi}}\right) \nonumber \\
&=&\text{Pf}(\mathbb{X}),
\label{norm_overlap}
\end{eqnarray}
where a convention in Ref.\cite{OM11} is employed for Grassmann vectors in this expression.
This convention is different from  Ref.\cite{Rob09} 
in the ordering of Grassmann numbers in Eq. (\ref{orderofxi}).
With this convention, a matrix $\mathbb{X}$ is given as
\begin{equation}
  \mathbb{X} = \left(\begin{array}{cc}
      Z^{(1)} & -\Lambda \\
      \Lambda & -\Lambda Z^{(0)*}\Lambda
    \end{array}\right),
  \label{Xmatrix}
\end{equation}
where  $\Lambda$ is defined as
\begin{equation}
  \Lambda_{ij} = \delta_{i+j,M+1}.
 \end{equation}
Thanks to the Pfaffian in the formula, 
an ambiguity in the sign assignment is removed, which was the primary concern in the Onishi formula
having an expression of a square root of a determinant.  
Note that, by Eq.(\ref{Grassmann_pf}) in the next section, we can immediately obtain the Pfaffian
form for the norm-overlap kernel Eq.(\ref{norm_overlap}). 

As for matrix elements with respect to one-quasiparticle HFB states,
we have derived a formula in Ref. \cite{OM11}.
With this formula, quantities such as
$\langle \Phi^{(0)} |c_{i}c^{\dagger}_{j}| \Phi^{(1)}  
\rangle$,  $\langle \Phi^{(0)} |c^{\dagger}_{i}c^{\dagger}_{j}| \Phi^{(1)}  
\rangle$, and  $\langle \Phi^{(0)} |c_{i}c_{j}| \Phi^{(1)}  
\rangle$, can be evaluated.
In our proof,  a variable transformation $\bar{\eta}=R\bar{\xi}$ is introduced.
A factorization of the anti-symmetric matrix $\mathbb{X}$ is carried out with a non-singular matrix $R$, 
as
\begin{equation}
 \mathbb{X}  = R^t \mathbb{J} R,
 \end{equation}
so as to derive the formula.
The matrix $\mathbb{J}$ in the above expression is given as
 \begin{equation}
  \mathbb{J} = \left(\begin{array}{cc}
      0 & I \\
      -I & 0\\
    \end{array}\right),
 \end{equation}
where $I$ is the unit matrix with the $M$ dimension.
The resultant formula is given in terms of $\mathbb{X}^{-1}$, as
\begin{eqnarray}
\langle \Phi^{(0)} |d_j d_i| \Phi^{(1)}\rangle &=&
\langle \Phi^{(0)} | \Phi^{(1)}\rangle (\mathbb{X}^{(-1)})_{ij}, \nonumber \\
& = & \text{Pf}(\mathbb{X})(\mathbb{X}^{(-1)})_{ij}.
\label{oneqmat}
\end{eqnarray}
The formula shown in Ref.\cite{OM11} contains extra matrix elements consisting of
the $U$ and $V$ factors, but in the above expression we show only the essential part
so as to be consistent with the generalization to multi-quasiparticle states, to be discussed below.

In order to derive a new formula for general matrix elements, we consider 
a generating function with Grassmann numbers in the subsequent sections. 
 
\section{Grassmann Integrals}

In this section, we review the Pfaffian type and determinant type of Grassmann integrals.
By cosidering these two types by comparison, it is possible to understand 
how natural the Pfaffian is in handling the HFB theory in a view of mathematical structure.
Based on the Refs. \cite{ZJ02,AA04},
we emphasize that the Pfaffian is mathematically more fundamental, and that
it plays an important role in the calculation of the matrix elements in the HFB theory.
Here, we use $\theta$'s and  $z$'s instead of $\xi$'s, for the sake of generality.

Firstly, the Pfaffian-type Grassmann integral is given by
\begin{equation}
 \int d\theta_{2n} \cdots d\theta_1
 \exp\left(\frac{1}{2}\theta^t A \theta\right)=Pf(A),
 \label{Grassmann_pf}
\end{equation}
where $\theta_1, \theta_2, \cdots, \theta_{2n}$ are Grassmann variables and 
$\theta^t=\left(\theta_1, \theta_2, \cdots, \theta_{2n}\right)$ is a Grassmann vector.
Matrix $A$ is a $2n \times 2n$ skew-symmetric matrix.
It can be easily proved by expanding the exponent, 
as shown in Refs.\cite{ZJ02,AA04}, for instance.

The determinant type of Grassmann integral is given by
\begin{equation}
 \int \prod_{\alpha}dz_\alpha^* dz_\alpha   
 \exp\left(-z^{*t} B z\right)=det(B)
 \label{Grassmann_det}
\end{equation}
where 
$z_1, z_2, \cdots, z_{n}$ are Grassmann variables and 
$z^*_i$ is a conjugate Grassmann number of $z_i$.
The $z^t=\left(z_1, z_2, \cdots, z_{n}\right)$ is a Grassmann vector and 
and $z^*$ is a conjugate vector of $z$. 
The $B$ is an $n \times n$ matrix.
Its proof is given, for instance, in Ref.\cite{NO85}.

Depending on how to define a Grassmann vector and how to set a form of the integrand, 
the outcome of the integral changes: it takes a form of either a determinant or the Pfaffian. 
It is worth noting, however, that the determinat type of Grassmann integral is included as a special case
of the Pfaffian-type Grassmann integral \cite{AA04}.
%
To demonstrate this fact, let us consider
a skew-symmetric matrix $A$ in the following form.
\begin{equation}
A=\left(\begin{array}{cc}
      0 & -B\Lambda\\
     (B\Lambda)^t & 0
    \end{array}\right),
\end{equation}
where $B$ is the matrix appearing in Eq.(\ref{Grassmann_det}).
In the Grassmann vector $\theta$ in Eq.(\ref{Grassmann_pf}), 
we choose $\theta_i=z_i^*$ for $i=1,\cdots, n$ and  $\theta_i=z_{2n+1-i}$ for $i=n+1,\cdots, 2n$, 
namely, a bipartite representation $\theta^t=(z^*,\Lambda z)^t$ is employed here.
Then, Eq.(\ref{Grassmann_pf}) reduces to  Eq.(\ref{Grassmann_det})
because of 
\begin{equation}
Pf(A)=(-1)^{n}det(B),
\end{equation}
and
\begin{equation}
d\theta_{2n} \cdots d\theta_1 = (-1)^{n}\prod_{\alpha=1}^{n}dz_\alpha^* dz_\alpha.
\end{equation}
From this result, it is possible to regard that the Pfaffian-type Grassmann integral is more fundamental. 
We call this property the determinant-Pfaffian correspondence.

The Grassmann integrals and  Wick's theorem are closely related to each other. 
It is known that a technique of generating functions is useful for finding a relation between them \cite{NO85}.
By making use of the determinant-type Grassmann integral Eq.(\ref{Grassmann_det}),
matrix element $\left< z_{i_1}\cdots z_{i_m}z^{*}_{i_m}\cdots z^{*}_{i_1} \right>$ 
can be expressed in a compact way as
\begin{eqnarray}
&& \left< z_{i_1}\cdots z_{i_m}z^{*}_{i_m}\cdots z^{*}_{i_1} \right> \nonumber \\
&=&\frac
{ \int \prod_{\alpha}dz_\alpha^* dz_\alpha   
 \exp\left(\frac{1}{2}z^{*t} B z\right)z_{i_1}\cdots z_{i_m}z^{*}_{i_m}\cdots z^{*}_{i_1}.
 }
 {\int \prod_{\alpha}dz_\alpha^* dz_\alpha   
 \exp\left(\frac{1}{2}z^{*t} B z\right)} \nonumber \\
 &=&det((B^{-1})_I),
 \label{Grassmann_det_wick}
\end{eqnarray}
with $m \le n$.
This relation can be obtained with a generating function proposed in Ref.\cite{NO85}.
As introduced in Sec.II, a compact notation for a group of indices\cite{IW00} is employed here again.
An $m \times m$ sub-matrix $(B^{-1})_I$ with $I=\{i_1, i_2, \cdots ,i_{m}\}$, is defined as
\begin{equation}
((B^{-1})_I)_{k,l}=(B^{-1})_{{i_k},{i_l}}
\end{equation}
for all $k$ and $l$ $\in I$.
From a viewpoint of Wick's theorem,
this relation implies that the matrix element $\left< z_{i_1}\cdots z_{i_m}z^{*}_{i_m}\cdots z^{*}_{i_1} \right>$ 
can be rewritten in terms of the contractions $\left< z_i z^{*}_j \right>$'s because of 
$(B^{-1})_{i,j}=\left< z_i z^{*}_j \right>$.
This result is general and needs no assumption of specific systems for the derivation,
which is presented in Eqs.(2.84)-(2-87) of Ref.\cite{NO85} through a generating function.

On the contrary, for the evaluation of the HFB matrix elements Eq.(\ref{qpmat_grassmann}), 
the Pfaffian version of Eq.(\ref{Grassmann_det_wick}) is useful. 
In a similar manner to the determinant case, we expect that
$\left< \bar{\xi}_{i_{2m}} \cdots \bar{\xi}_{i_{1}} \right>$ should be decomposed into combinations of
the contractions $\left< \bar{\xi}_i\bar{\xi}_j \right>$'s.
The existence of this relation is likely due to the determinant-Pfaffian correspondence. 
Its derivation can, in fact,  be easily carried out in the same procedure
with a generating function, which will be discussed in the next section in detail.
We learned that this Pfaffian relation is briefly mentioned in Ref.\cite{ZJ02},
and it is summarized in appendix A of a very recent paper \cite{CSS11} from a mathematical point of view.
However, in the context of a physical application of the relation,
we are the first to present the relation in detail how to evaluate the matrix elements.
In addition, a comparison of the result to the conventional expansion by the usual Wick theorem
is demonstrated for the first time in this paper.

\section{Generating function} 

To evaluate general matrix elements for multi-quasiparticle states, 
we introduce a generating function defined by
\begin{equation}
G(\bar{J})\equiv 
\frac
{
  \int \mathcal{D}\bar{\xi}
 \exp\left(
    \frac{1}{2}\bar{\xi^t}\mathbb{X}\bar{\xi}
   +\frac{1}{2}\bar{J^t}\bar{\xi}-\frac{1}{2}\bar{\xi^t}\bar{J}
 \right)
}
{
  \int \mathcal{D}\bar{\xi}
  \exp\left(\frac{1}{2}\bar{\xi^t}\mathbb{X}\bar{\xi}\right)
}
\label{generating}
\end{equation}
where a ``source'' $\bar{J}$ is defined as a Grassmann vector
\begin{equation}
  \bar{J}^t \equiv (J_1^*,J_2^*,\cdots,J_M^*,
  J_M,J_{M-1},\cdots,J_1).
  \label{vector_J}
\end{equation}

Introducing a new variable $\bar{\xi'}$ by shifting $\bar{\xi}$ as
\begin{equation}
  \bar{\xi'} \equiv  \bar{\xi}-\mathbb{X}^{-1}\bar{J}, 
\end{equation}
the following relation is obtained
\begin{equation}
\frac{1}{2}\bar{\xi^t}\mathbb{X}\bar{\xi}
+\frac{1}{2}\bar{J^t}\bar{\xi}-\frac{1}{2}\bar{\xi^t}\bar{J}= 
\frac{1}{2}\bar{\xi'^t}\mathbb{X}\bar{\xi'}+
 \frac{1}{2}\bar{J^t}\mathbb{X}^{-1}\bar{J}.
\end{equation}
The generating function thus becomes
\begin{equation}
G(\bar{J})=
\exp\left( \frac{1}{2}\bar{J^t}\mathbb{X}^{-1}\bar{J}\right).
\label{generating2}
\end{equation}

A differentiation of the original form of the generating function Eq.(\ref{generating})
with respect to the source vector $\bar{J}$,
gives rise to the matrix element $\mathcal{M}_I$ in Eq.(\ref{qpmat_grassmann}) in the form of
\begin{eqnarray}
& &\frac{\delta^{(2m)}}{\delta J_{i_{2m}}\cdots\delta J_{i_1}} G(\bar{J})
\Bigr|_{\bar{J}=0}  \nonumber \\
&=&
\frac
{
  \int \mathcal{D}\bar{\xi}
  \exp\left(
    \frac{1}{2}\bar{\xi^t}\mathbb{X}\bar{\xi}
   +\frac{1}{2}\bar{J^t}\bar{\xi}-\frac{1}{2}\bar{\xi^t}\bar{J} \right)
   \bar{\xi}_{i_{2m}}\cdots\bar{\xi}_{i_{1}}
}
{
  \int \mathcal{D}\bar{\xi}
\exp\left(\frac{1}{2}\bar{\xi^t}\mathbb{X}\bar{\xi}\right)
}\Bigr|_{\bar{J}=0} \nonumber \\
&=& 
\frac{
  \int \mathcal{D}\bar{\xi}
 \exp\left(\frac{1}{2}\bar{\xi^t}\mathbb{X}\bar{\xi}\right)
 \bar{\xi}_{i_{2m}}\cdots\bar{\xi}_{i_{1}}
}
{
\int \mathcal{D}\bar{\xi}
\exp\left(\frac{1}{2}\bar{\xi^t}\mathbb{X}\bar{\xi}\right)
}.
\label{generating3}
\end{eqnarray}

On the other hand, by differentiating the form of Eq.(\ref{generating2}),
we can obtain an alternative expression for
$\frac{\delta^{(2m)}}{\delta J_{i_{2m}}\cdots\delta J_{i_1}} G(\bar{J})
\Bigr|_{\bar{J}=0}$.
For the aim to express in a compact manner, we introduce
a $2m$-dimensional sub-vector $\bar{J}_I$ and a $2m \times 2m$ sub-matrix $A_I$,
with $I=\{i_1, i_2, \cdots ,i_{2m}\}$. They are defined as
\begin{eqnarray}
(\bar{J}_I)_k&=&\bar{J}_{i_k}  \nonumber \\
(A_I)_{k,l}&=&A_{{i_k},{i_l}}
\end{eqnarray}
for all $k$ and $l$ $\in I$.
As the Grassmann numbers in a subgroup $[2M]-I$ vanishes, the derivative of the generating function
is calculated to be
\begin{eqnarray}
& &\frac{\delta^{(2m)}}{\delta J_{i_{2m}}\cdots\delta J_{i_1}} G(\bar{J})\Bigr|_{J=0}  \nonumber \\
&=&\frac{\delta^{(2m)}}{\delta J_{i_{2m}}\cdots\delta J_{i_1}}
\exp\left( \frac{1}{2}\bar{J_I}^t(\mathbb{X}^{-1})_I \bar{J}_I\right)  \nonumber \\
&=& \text{Pf}((\mathbb{X}^{-1})_I).
\label{generating4}
\end{eqnarray}
In the above calculation, we use a useful relation for any skew-symmetric matrix $A$ with $2m \times 2m$ dimension
and $2m$ Grassmann numbers $z$'s, which is
\begin{eqnarray}
&&\frac{\delta^{(2m)}} {\delta z_{2m} \cdots\delta z_1}  
\exp\left( \mathbf{A} \right)  \nonumber \\
&=&\frac{\delta^{(2m)}} {\delta z_{2m} \cdots\delta z_1} 
\frac{1}{m!}\mathbf{A}^{m}   \nonumber \\
&=&\frac{\delta^{(2m)}} {\delta z_{2m} \cdots\delta z_1}  
\text{Pf}(A)z_1 z_2 \cdots z_{2m}   \nonumber \\
&=& \text{Pf}(A),
\label{pfaff_derivative}
\end{eqnarray}
where $\mathbf{A}=\frac{1}{2} \sum_{i,j=1}^{2m} z_i A_{i,j} z_j$.
Eq.(\ref{zn}) in Appendix is also used for the above derivation.

Equating Eqs.(\ref{generating3}) and (\ref{generating4}),
a new formula is obtained for matrix elements between different multi-quasiparticle states:
\begin{eqnarray}
 \mathcal{M}_I &=& 
 \langle \Psi^{(0)} | \Psi^{(1)}\rangle \text{Pf}((\mathbb{X}^{-1})_I), \nonumber  \\
 & = & \text{Pf}(\mathbb{X}) \text{Pf}((\mathbb{X}^{-1})_I).
  \label{qp_formula}
\end{eqnarray}
This is a generalization of Eq.(\ref{oneqmat})
because $\text{Pf}(A)=a$ as in Eq.(\ref{pfaff22}) for a $2\times 2$ anti-symmetric matrix 
$A= \left(\begin{array}{cc}
      0 & a \\
      -a & 0\\
    \end{array}\right)$.  
It should be noted here that we can also derive Eq.(\ref{Grassmann_det_wick}) from Eq.(\ref{qp_formula}) by 
the determinant-Pfaffian correspondence, which is to be presented in Section IV.

As we will discuss below, our formula turns out 
to be an extension of the traditional generalized Wick theorem\cite{RS80}.

\section{Generalized Wick's theorem}
The standard approach to calculate matrix elements of many-body operators is
to use  generalized Wick's theorem \cite{RS80}, in which a chain of operators is expanded 
in terms of normal orders and contractions.

Contractions in the HFB theory correspond to the transition densities 
($\rho^{(01)}$ and  $\tilde{\rho}^{(01)}$), and the transition pairing tensors
($\kappa^{(01)}$  and $\kappa^{(10)*}$). 
These quantities are mathematically expressed as 
\footnote{In Ref.\cite{RS80}, 
$\rho^{(01)}$ is employed as the transition density because $\tilde{\rho}^{(01)}=1-\rho^{(01)}$, 
but we use  $\tilde{\rho}^{(01)}$ in the present work.}
\begin{eqnarray}
\rho^{(01)}_{i,j} &=&
\frac{
 \langle \Phi^{(0)}|c_{j}^\dag c_i |\Phi^{(1)}\rangle 
}
{
 \langle \Phi^{(0)}|\Phi^{(1)}\rangle 
}, \nonumber \\
\tilde{\rho}^{(01)}_{i,j} &=&
\frac{
 \langle \Phi^{(0)}|c_{i} c_j^\dag |\Phi^{(1)}\rangle 
}
{
 \langle \Phi^{(0)}|\Phi^{(1)}\rangle 
}, \nonumber \\
\kappa^{(01)}_{i,j} &=&
\frac{
 \langle \Phi^{(0)}|c_{j} c_i |\Phi^{(1)}\rangle 
}
{
 \langle \Phi^{(0)}|\Phi^{(1)}\rangle 
}, \nonumber \\
\kappa^{(10)*}_{i,j} &=&
\frac{
 \langle \Phi^{(0)}|c_{i}^\dag c_j^\dag |\Phi^{(1)}\rangle 
}
{
 \langle \Phi^{(0)}|\Phi^{(1)}\rangle
},
\end{eqnarray}
where $1\le i,j \le M$. 
In accordance with generalized Wick's theorem, 
any matrix elements in Eq.(\ref{m_qp}) can be expressed by these contractions in a combinatorial way. 

In the present method, these basic contractions are connected to $\mathbb{X}^{-1}$ directly,
for instance, 
\begin{eqnarray}
\kappa^{(10)*}_{i,j} &=&
\frac{
-\langle \Phi^{(0)}|d_{j}^\dag d_i^\dag |\Phi^{(1)}\rangle 
}
{
 \langle \Phi^{(0)}|\Phi^{(1)}\rangle
} \nonumber \\
&=& -(\mathbb{X}^{-1})_{ij},
\end{eqnarray}
where Eq.(\ref{oneqmat}) is used. 
In general, a relation between $\mathbb{X}^{-1}, \rho$, and $\kappa$ is given as
\begin{eqnarray}
\mathbb{X}^{-1}&=&\left( 
\begin{array}{cc}
-\kappa^{(10)*}& \tilde{\rho}^{(01)t} \\
-\tilde{\rho}^{(01)} & \kappa^{(01)} \\
\end{array} 
\right), \nonumber \\
&=&\left( 
\begin{array}{cc}
-\kappa^{(10)*} & (1-\rho^{(01)})^t \\
\rho^{(01)}-1  & \kappa^{(01)} \\
\end{array} 
\right).
\label{Xmatrix-density}
\end{eqnarray} 

The transition densities and transition pairing tensors can be also expressed in terms of the
skew-symmetric matrices $Z^{(p)}$ in the Thouless form Eq.(\ref{Thouless}),
as shown in Appendix E of Ref.\cite{RS80}, which needs somewhat lengthy calculations for the proof.
In the present method, however, $\mathbb{X}^{-1}$  is directly calculated from Eq.(\ref{Xmatrix}), 
which results in
\begin{eqnarray}
&&\left( 
\begin{array}{cc}
 1 &  0 \\ 
 0  &  \Lambda \\
\end{array} 
\right)
\mathbb{X}^{-1}
\left( 
\begin{array}{cc}
 1 &  0 \\ 
 0  &  \Lambda \\
\end{array} 
\right)\\
&=&
\left( 
\begin{array}{cc}
  Z^{*(0)}{(Z^{(1)} {Z^{*(0)}}-1)}^{-1} &          {(1-Z^{*(0)} {Z^{(1)}})}^{-1}  \\ 
            {(Z^{(1)} {Z^{*(0)}}-1)}^{-1}  &  Z^{(1)}{(1-Z^{*(0)} {Z^{(1)}})}^{-1}  \\
\end{array} 
\right). \nonumber
\end{eqnarray} 
This result is consistent with the expressions (E.54) in Ref.\cite{RS80}.

Let us now confirm that our formula to calculate many-body matrix elements
is consistent with the result obtained with generalized Wick's theorem.
As an example, let us consider a case of two-quasiparticle states 
$|\Phi_{k_1k_2}^{(p)}\rangle=\beta_{k_1}^{\dag}\beta_{k_2}^{\dag}|\Phi^{(p)}\rangle$.
In the evaluation of the relevant norm-overlap, several types of many-body matrix elements need to be
calculated, say ${\langle \Phi^{(0)}|c_{l_4} c_{l_3} c_{l_2}^\dagger c_{l_1}^\dagger
|\Phi^{(1)}\rangle }$.
This quantity can be calculated by means of our formula Eq.(\ref{qp_formula}) as,
\begin{eqnarray}
&& \frac{\langle \Phi^{(0)}|c_{l_4} c_{l_3} c_{l_2}^\dagger c_{l_1}^\dagger
|\Phi^{(1)}\rangle }
{\langle \Phi^{(0)}|\Phi^{(1)}\rangle} \\ \nonumber
&=& \text{Pf}((\mathbb{X}^{-1})_{ \{ l_1, l_2, \bar{l_3}, \bar{l_4} \}})
 \\ \nonumber
&=&
 (\mathbb{X}^{-1})_{l_1,l_2}(\mathbb{X}^{-1})_{\bar{l}_3,\bar{l}_4}
 - (\mathbb{X}^{-1})_{l_1,\bar{l}_3}(\mathbb{X}^{-1})_{l_2,\bar{l}_4}\\ \nonumber
&& +(\mathbb{X}^{-1})_{l_1,\bar{l}_4}(\mathbb{X}^{-1})_{l_2,\bar{l}_3}. \\ \nonumber
&=&-\kappa_{l_1,l_2}^{(10)*} \kappa_{l_3,l_4}^{(01)}
-\tilde{\rho}_{l_3,l_1}^{(01)}\tilde{\rho}_{l_4,l_2}^{(01)}
+\tilde{\rho}_{l_4,l_1}^{(01)}\tilde{\rho}_{l_3,l_2}^{(01)},
\label{m_2qp}
\end{eqnarray} 
where $0 \le l_1, l_2, l_3, l_4\le M$. 
In the above equations, the second line is obtained with the formula Eq.(\ref{qp_formula}),
and the third line is an explicit expansion of the Pfaffian due to the definition Eq.(\ref{pfaff44}).
The last line is a substitution with the result in Eq.(\ref{Xmatrix-density}).
The last line is consistent with the result derived from traditional Wick's theorem.
In other words, the present Pfaffian formula is an alternative representation of Wick's theorem.
For details of generalized Wick's theorem, see Ref.\cite{OY66} for instance. 


Recently, Perez-Martin and Robledo also reformulated 
Wick's theorem based on the statistical mechanics (Gaudin's theorem) 
to give  a closed expression in Ref. \cite{Rob07}.
Also, Bertsch and Robledo presented another closed form with the Pfaffian\cite{BR11}. 
In comparison with these formulae, 
the present formula has a fairly compact and simple expression.
The main reason for this compactness and simplicity comes from an introduction of $\mathbb{X}^{-1}$, by which the transition density and transition pairing tensor can be unified.
As a consequence,
the expression of $\mathcal{M}_I$ 
becomes free from combinatorial complexity in applying generalized Wick's theorem.

\section{Balian-Brezin conjecture}

In nuclear structure physics, generalized Wick's theorem is quite important especially 
for a description of multi-quasiparticle states based on an HFB vacuum. Here we revisit the contents of
the theorem  historically.

In Ref.\cite{LO55}, L\"{o}wdin first noticed that an identity for determinants, 
called  the Lewis Carroll (Dodgson) formula  or Desnanot-Jacobi adjoint matrix theorem\cite{DB99}, 
can be applied to the density matrix for Slater determinants\cite{BB69}. 
In Ref.\cite{BB69}, Balian and Brezin conjectured that a Pfaffian counterpart of Sylvester's identity\cite{Syl} for Slater determinants would be useful for the HFB theory.
However, there has been no progress in this conjecture since its proposal in 1969
as shown in the first footnote of Ref.\cite{AB12}.
In this section, we discuss this conjecture.

To approach the Balian-Brezin conjecture, we consider the Ref.\cite{AB12} where 
they consider an another way to carry out the Grassmann integrations in Eq.(\ref{qpmat_grassmann}).
The Grassmann variables $\{\xi_1^*,\xi_2^*,\cdots,\xi_M^*,\xi_M,\xi_{M-1},\cdots,\xi_1\}$
are divided into two groups.
The first group of the Grassmann variables consists of  $I=\{i_1,\cdots, i_{2m}\}$, which corresponds to
the indices appearing in the operators $d_i$ in Eq.(\ref{m_qp}).
The other group consists of the rest of the indices in the total model space, 
which is denoted as $\bar{I}=[2M]-I$ meaning the complimentary group of $I$.

Due to a reordering of the Grassmann variables,
the integral measure is rewritten as  
\begin{equation}
\prod_{\alpha}d\xi_\alpha^* d\xi_\alpha=
\mathcal{D}\bar{\xi}_I
\mathcal{D}\bar{\xi}_{\bar{I}}(-)^{|I|}
\end{equation}
where $ |I|=\sum_{k=1}^{2m} i_k $.
Here  
$\mathcal{D}\bar{\xi}_I=d\bar{\xi}_{i_1} \cdots d\bar{\xi}_{i_{2m}}$
and 
$\mathcal{D}\bar{\xi}_{\bar{I}}
=d\bar{\xi}_{j_{2M-2m}} \cdots d\bar{\xi}_{j_1}, ( j_1 < j_2 < \cdots < j_{2M-2m}, j_k \in \bar{I}$).
Integration over $\mathcal{D}\bar{\xi}_I$ is easily carried out and gives rise to unity.
The matrix element for $\mathcal{M}_I$ thus becomes
\begin{eqnarray}
\mathcal{M}_I&=&(-)^{|I|}\int \mathcal{D}\bar{\xi}_{\bar{I}}
 \exp\left({\frac{1}{2}{\bar{\xi}_{\bar{I}}}^t\mathbb{X}_{\bar{I}}\bar{\xi}_{\bar{I}}}\right) \nonumber \\
  &=& (-)^{|I|}\text{Pf}(X_{\bar{I}}).
\label{avezbender}
\end{eqnarray}
Here, we use the notation $A_{\bar{I}}$ for a $2(M-m) \times 2(M-m)$ sub-matrix of a matrix $A$,
which is produced by removing the matrix elements $A_{i,j}$ for $i,j=i_1,\cdots, i_{2m}$ from the original matrix $A$.
Note that the resultant formula Eq.(\ref{avezbender}) is nothing to do with generalized Wick's theorem, 
unlike the formula we have derived in Eq.(\ref{qp_formula}).

Nonetheless, it is possible to relate our result with the above result, 
through an identity for the Pfaffian. 
By comparing Eq.(\ref{qp_formula}) to Eq.(\ref{avezbender}), there should be a relation
\begin{equation}
 \text{Pf}(\mathbb{X}) \text{Pf}((\mathbb{X}^{-1})_I)=(-)^{|I|} \text{Pf}(\mathbb{X}_{\bar{I}}).
 \label{pfcarrol}
\end{equation}
We found that this relation holds for any skew-symmetric matrix $\mathbb{X}$, 
and that this Pfaffian identity has been recently discovered by mathematicians\cite{IW00}. 
Their proof is, however, given in an elementary way without relying on any Grassmann integrations \cite{IW00,CSS11a}.
This identity is called the Pfaffian version of  the Lewis Carroll 
(Dodgson) formula \cite{IW00}, whose original version holds for determinant.  

Through the comparison between our result and the one obtained by Avez and Bender \cite{AB12},
we found that the both Pfaffian formula are firmly connected by the Pfaffian identity,
which is just what Balian and Brezin conjectured for the HFB case\cite{BB69}.


\section{Summary} 

In the present paper, we presented a compact and closed formula 
to evaluate matrix elements for multi-quasiparticle HFB states, and 
a relation to generalized Wick's theorem is discussed.

To calculate multi-quasiparticle matrix elements, 
we started our derivation from a fact that the Thouless 
form of the HFB wave function can be rewritten by the Fermion 
coherent state with Grassmann numbers,
by following Robledo's approach \cite{Rob09}.
By a generating function with Grassmann numbers, 
we could derive a new Pfaffian formula to calculate matrix elements between 
multi-quasiparticle HFB states.
This formula is an extension of the norm-overlap kernels of one-quasiparticle states
which we presented in Ref.\cite{OM11}. 
In traditional generalized Wick's theorem, the transitional density and pairing tensor are
utilized in a combinatorial way, while, in this new formula, these basic contractions are 
utilized with no distinction. In consequence, the formula becomes simple and compact 
and is free from combinatorial complexity.
  
Besides, by comparing our Pfaffian formula to the one obtained by Avez and Bender\cite{AB12},
we found that the both Pfaffian formulae are closely related to the Pfaffian version 
of the Lewis Carroll formula in pure mathematics\cite{IW00}. 
In Ref.\cite{BB69}, Balian and Brezin conjectured a usefulness of extended Sylvester's identity 
by means of the Pfaffian \cite{Syl}. 
The present paper gives an affirmative answer to their conjecture.

\section*{Appendix}
For a skew-symmetric matrix $A$ with dimension $2n\times 2n$,
whose matrix elements are $a_{ij}$, the Pfaffian is defined as   
\begin{equation}
\text{Pf}(A)\displaystyle \equiv\frac{1}{2^{n}n!}\sum_{\sigma\in S_{2n}}
{\rm sgn}(\sigma)\prod_{i=1}^{n}a_{\sigma(2i-1)\sigma(2i)}
\label{defpfaff}
\end{equation}
where $\sigma$ is a permutation of $\{1,2,3,\cdots , 2n\}$, ${\rm sgn}(\sigma)$ is its sign,
and $S_{2n}$ represents a symmetry group.

For Grassmann numbers $z_1, z_2, \cdots, z_{2n}$, if we define $Z$ as 
\begin{equation}
Z=\sum_{i<j} a_{ij}z_i z_j,
\end{equation}
then $Z^n$ is shown by the Pfaffian as 
\begin{equation}
\frac{1}{n!}Z^n=\text{Pf}(A)z_1 z_2 \cdots z_{2n}.
\label{zn}
\end{equation}
This is a definition of the Pfaffian in the exterior algebra.

According to Eq.(\ref{defpfaff}), for a $n \times n$ $(n=\text{odd})$ matrix,
$\text{Pf}(A) =0$.

For a $2 \times 2$ matrix, 
\begin{equation}
\text{Pf}(A) =a_{12}.
\label{pfaff22}
\end{equation}

For a $4 \times 4$ matrix, 
\begin{equation}
\text{Pf}(A) =a_{12}a_{34}-a_{13}a_{24}+a_{14}a_{23}.
\label{pfaff44}
\end{equation}


\begin{thebibliography}{99}
\bibitem{Syl} In Ref.\cite{BB69}, there is a phrase of 'an extension to Pfaffians of Sylvester's identity' but the Lewis Carroll formula or Desnanot-Jacobi adjoint matrix theorem\cite{DB99} is more appropriate than Sylvester's identity.
\bibitem{CSS11a}  We note that the same proof based on the Grassmann integration has been pointed out in appendix A of the Ref. \cite{CSS11} in a view of a mathematical interest.
\end{thebibliography}
\end{document}